\documentclass[journal, twocolumn]{IEEEtran}

\usepackage{graphicx}
\graphicspath{{./figures/}}
\DeclareGraphicsExtensions{.pdf}
\usepackage{epstopdf}
\usepackage{subcaption}
\usepackage{multirow}
\usepackage{amsmath}
\usepackage[numbers]{natbib}
\usepackage{url}

\begin{document}

\title{Application of the NIP wave theorem to PSDM and an approximation for RPSM to zero offset}

\author{Joerg F. Schneider \\
  Bureau of applied Geophysics, Nordstemmen, Germany}

\maketitle

\begin{abstract}

The concept of residual migration to zero offset is introduced for the case that a prestack migration and a subsequent residual moveout analysis have been performed for a seismic survey and a new depth model has not yet been determined. Travel times of reflected events for individual traces are determined for diffraction points situated along horizons of analysis. These events are downward continued into the model used for the migration of the survey. The NIP wave theorem can then be applied to events exhibiting small subsurface offsets in order to determine residual radii of curvature to be used in the updating of seismic velocity model by seismic stripping. Alternatively, it is suggested to construct an aplanat and hence to determine a focus point of the reflected event without knowledge of the true velocity model. The distance of the estimated point of focus from the observed zero offset response of the migrated reflector serves as an indication of the focusing of the considered event at the particular subsurface offset and can be used as a measure for the necessity of a subsequent remigration of the survey with a new velocity model. A summation over the first Fresnel zone at the migrated zero offset position is suggested as an alternative to conventional stacking leading to improvement of the signal to noise ratio. Applications to model computations and to a seismic survey over an overthrust structure show promising results regarding the applicability of the suggested method.

\end{abstract}

\section{Introduction}

Prestack migrations (PSM) have been performed for decades for the processing of seismic decades either as prestack time (PSTM) or prestack depth (PSDM) migrations (Yilmaz, 2001, Bednar, 2005, Robein 2010, Al-Chalabi, 2014). Previously, but also in last years (Douma and de Hoop, 2006) the migration and demigration of interpreted reflection responses has been considered (Lambare et alii, 2008, Guillaume et alii, 2008).\\

Recently in (Schneider, 2017, 2020) a different approach was suggested: after PSM migrated reflection responses are estimated from residual moveout (RMO) analyses. The migrated responses were demigrated and migrated into a new velocity model by applying aplanatic constructions: according to Hagedoorn (1954) the migrated response can be obtained as the envelope of all surfaces of equal reflection times of the source receiver pairs. Mappings were used to assess the necessity of a new PSDM and a kinematic extension was presented for residual PSM (RPSM) over the first Fresnel zone (Born, 1985, Sheriff, 1980, Lindsey, 1989). Subsequently this approach was extended (Schneider, 2019) for a RPSM to zero offset in the domain described by the original velocity model.\\

In this contribution different concepts will be presented: only the velocity model used for the PSM will be considered, it will not be assumed that a new velocity model is present. The NIP wave theorem (Hubral and Krey, 1980) will be applied to events exhibiting small subsurface offsets in order to estimate residual radii of curvature with a view to update the velocity model.
An alternative approach will be introduced, where the focusing of a migrated response can be estimated from the curvatures of diffraction responses without knowledge of the correct velocity model. A summation over the first Fresnel zone is suggested as an alternative to stacking leading to an improvement of the signal to noise ratio. \\

 \section{Application of the NIP wave theorem to PSDM}

\begin{figure}
   \captionsetup{justification=justified, singlelinecheck=off}
\centering
\includegraphics[width=0.45\textwidth]{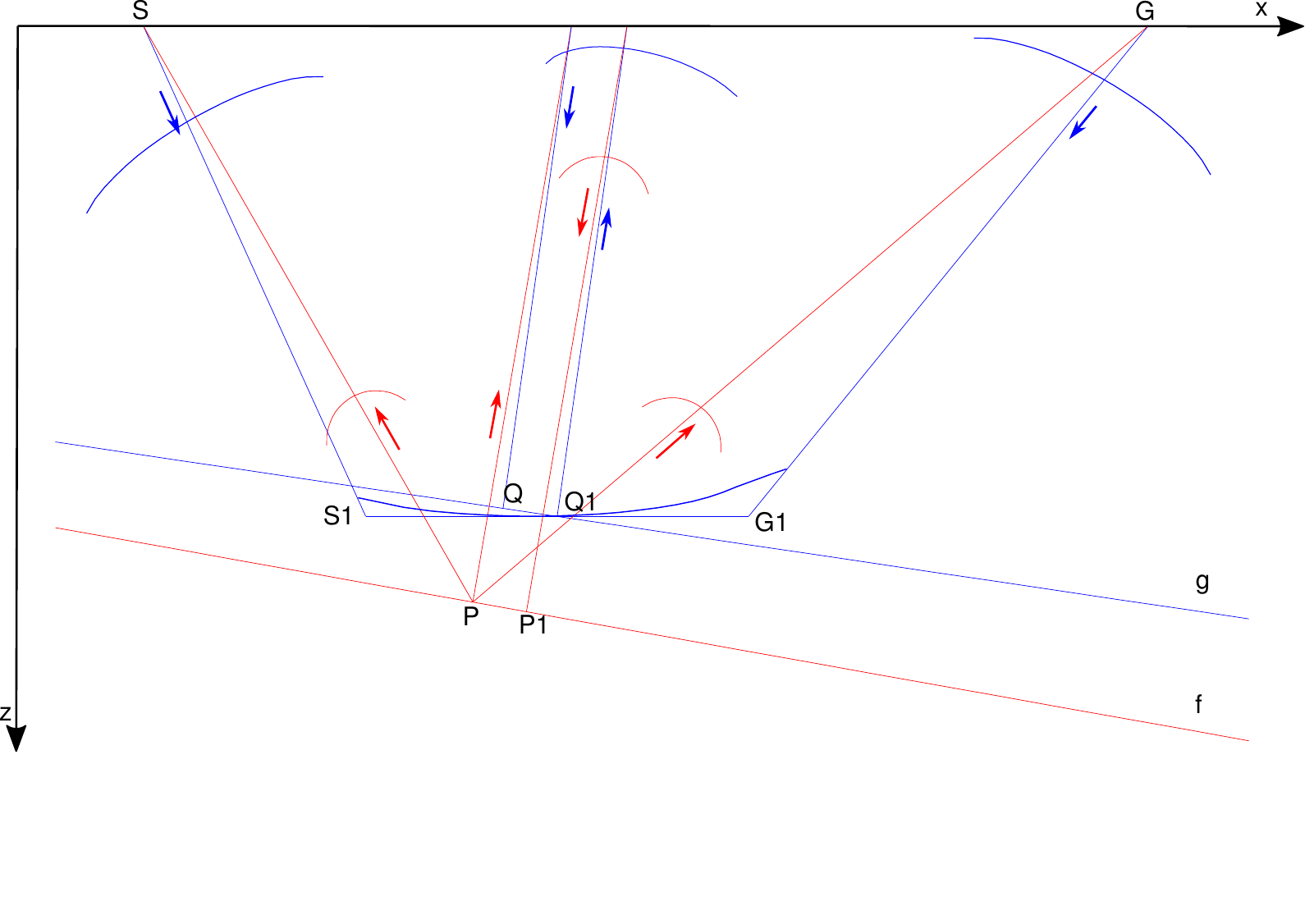}
\caption{Downward continuation of a source receiver pair and a corresponding NIP wave}
\label{fig:fig1}
\end{figure}

Figure~\ref{fig:fig1} shows a reflector f of constant dip in the depth domain in red (velocity $v_t$) and its migrated zero offset image g in blue where the migration velocity $v_m$ is too low. A reflected ray at P is shown for a shot and receiver S and G as well as the zero offset ray originating at the reflection point P and terminating at the point Q on g. All rays are downward continued into the medium with migration velocity $v_m$. It is then possible to find a point Q1 which bisects the downward continued source receiver on g at equal depth such that S1Q1=Q1G1 and to approximate the downward continued wavefront for this configuration at Q1  in terms of the radius of curvature of a zero offset ray according to:

\begin{equation}
  \label{rda}
t(x)=(sin(\gamma)/v_m)x_s+x_s^2(cos^2\gamma)/(2v_mR_0)
\end{equation}

 \begin{figure}
   \captionsetup{justification=justified, singlelinecheck=off}
\centering
\includegraphics[width=0.45\textwidth]{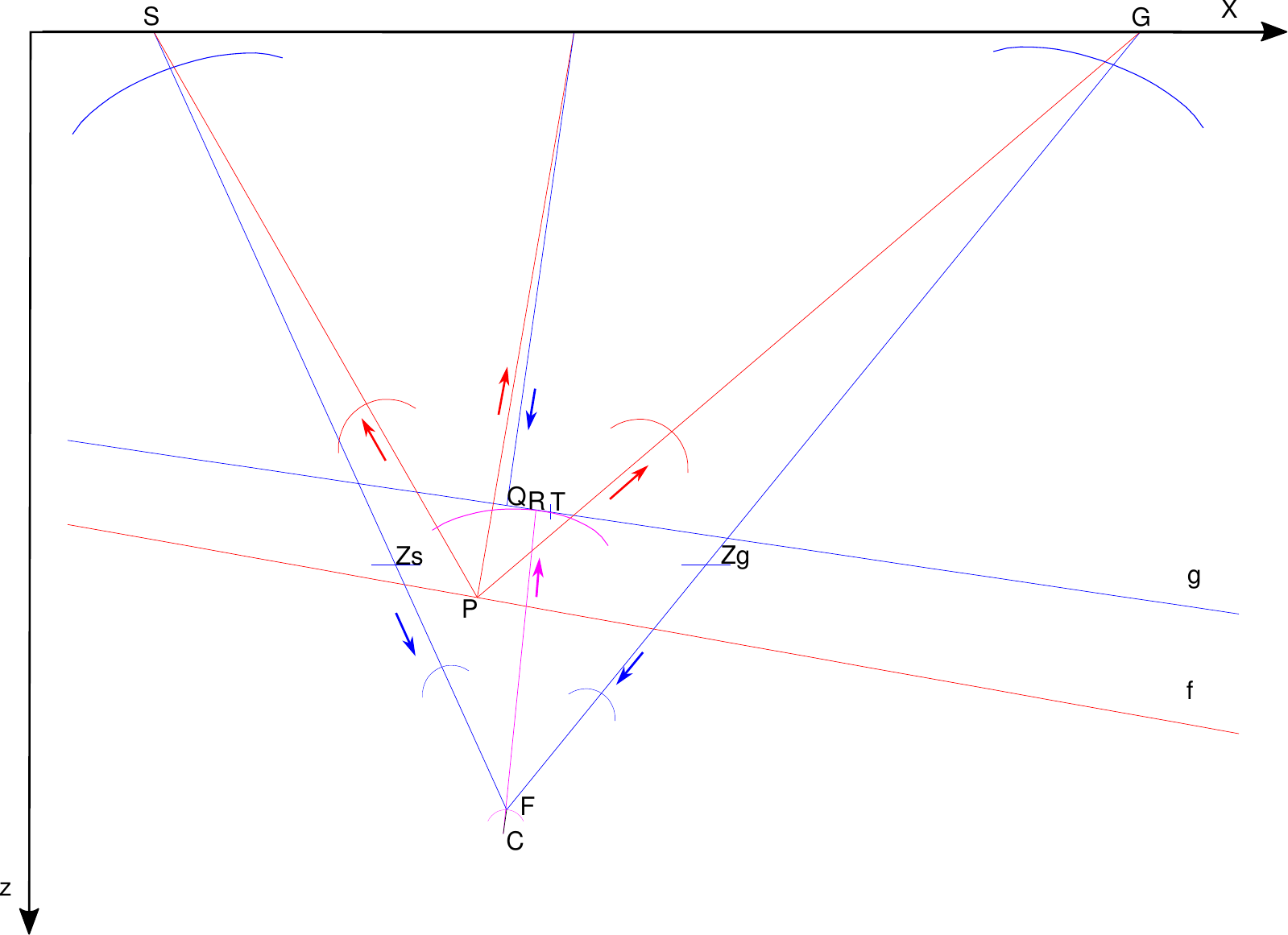}
\caption{Approximate RPSM to zero offset}
\label{fig:fig2}
\end{figure}

 where $\gamma$ is the dip angle of $g$, $x_s$ is the distance $S_1Q_1=Q_1G_1$ and $R_0$ is the residual radius of curvature of the downward continued zero offset ray at $Q_1$ (Figure~\ref{fig:fig1}), (Hubral and Krey, 1980). From (\ref{rda}) it is in principle possible to estimate $R_0$ from a residual moveout analysis after PSDM and to use this quantity for an update of the velocity model. However, it should be noted that
\begin{itemize}
\item
  the approximation in  (\ref{rda}) is only second order in $x_s$. Typically it is valid only for small subsurface offsets, corresponding to surface offsets up to only 1000m, even for depth models of moderate complexity. For larger source receiver distances short offset approximations have to be estimated, e.g. from higher order residual moveout analyses. On the other hand, e.g. in applications of seismic stripping, changes in the overburden due to curvatures of interfaces are fully accounted for.
\item
  a positional bias (to be explained in the next section) exists in the migrated domain, corresponding to a bias $PP_1$ which can only be estimated with a knowledge of the true velocity model. Computations for simple models show that this bias is often larger than the bias introduced by an erroneous migration according to the common offset or the downward continuation approach.
\end{itemize}

\section{An approximation for RPSM to zero offset}

Figure~\ref{fig:fig2} shows the same configuration as Figure~\ref{fig:fig1}. Again   expanding wavefronts from $P$ are considered with the difference that now the reflected event is fully migrated with subsurface offset defined by $Z_S$ and $Z_G$ (Dafni and Symes, 2018), and projected midpoint $T$ on $g$ (Figure~\ref{fig:fig2}). The positional bias for this event is the lateral distance between $T$ and the migrated zero offset ray at $Q$ on $g$. Now a technique which was used in (Schneider, 1989) is employed: the legs of the rays are further continued to the point of intersection $F$, with residual radii of curvature $R_S$ and $R_G$. From these quantities the radius $R_F$ can be determined by

\begin{equation}
  \label{rdb}
R_G^{-1}=R_S^{-1}+(2/cos(\epsilon))R_F^{-1}
\end{equation}

where $\epsilon$ is the angle between the bisector and the ray legs at $F$. $R_F$ in (\ref{rdb}) defines the radius of curvature of the aplanatic curve perpendicular to the bisector for which a residual prestack migration over the first Fresnel zone could be performed if the appropriate time samples are available, the center $C$ of the circle is the image point of the reflection point $P$ (to second order). It is then possible to perform a mapping of the migrated sample at  $Z_S$, $Z_G$ along the wavefront expanding along the zero offset ray from $C$ to $R$ on $g$ (Figure 2) by employing an heuristic approach as in (Schneider, 2019), viz:

\begin{equation}
  \label{rdc}
g(x_R,z_R)\propto\int_g u(x)D^{-1}[f(x,z)]dzdx
\end{equation}

of the primary migration  at g, and the half differentiator $D^{-1}[f(t)]=F^{-1}(|2\pi\omega|^{1/2}e^{-i(\pi/4)sg(\omega)}F(f(t)))$. (Bleistein et alii, 1987), where  $F(f(t))$ is the Fourier transform of $f(t)$  in the frequency domain $\omega$. The function $u(x)$ is the density ratio of the migrated to the imaged events on $g$, the summation is performed over the first Fresnel zone (see second remark below).\\\\
The positional bias, i.e. the distance from the migrated zero offset image Q on g in Figure~\ref{fig:fig2} is reduced for isotropic homogeneous media as $QR < QT$ and $x_Q<x_R<x_T$ (Figure~\ref{fig:fig2}). The following is an outline of a proof of this statement for the case illustrated in Figure~\ref{fig:fig2} ($v_m < v_t$, positive slope of $f$):\\
For T and Q we have that $x_T> x_Q$ because the migrated zero offset ray at the midpoint of $z_s$ and $z_g$ originates to the right of $P$ at horizon $f$. Also, $x_T> x_R$, because the angle between the normal to $g$ and the x-axis is larger than the angle between F and the midpoint of $Z_S$ and $Z_G$(i.e. the median in the triangle with corners $F$, $Z_G$, $Z_S$) and the x-axis; equally the angle between $FC$ and the  x-axis (i.e. the bisector between $Z_sF$ and $FZ_g$) is larger than the angle between the normal to $g$ and the x-axis. These three angles are of equal sign, the differences beteen them are $O[(\Delta v /v_t))]$, with $\Delta v =v_t-v_m$. We have that $x_R-x_T = O[(\Delta v /v_t))]$, wheras $x_T-x_Q = O[(\Delta v /v_t)^2)]$ and the proposition follows if $\Delta v$ is small enough; it also follows that the distance $RT$ is small in this case. 
\\\\
It should be noted that:
\begin{itemize}
\item
  in Figure~\ref{fig:fig2} $CR$ defines the distance from the center of the focused event and hence the width of the Fresnel zone of the mapping to zero offset, which defines the lateral resolution (Born, 1985) – the mapping is supposed to be an alternative to conventional stacking: there will be lateral movement and it is suspected that the summation in equation (\ref{rdc}) will improve the signal to noise ratio whereas the lateral resolution is decreased with increasing $CR$.
\item
  the mapping should be performed from the position of the determined focal point; however, the corresponding time sample would only be available if the migration has been accomplished according to the downward continuation approach for the two way wave equation. Alternatively, a suitable approximation to the focused event could be used, e.g. in the case that a migration has been performed according to the one way wave equation with the migrated sample after residual moveout and a demigration followed by a migration as described above. This approximation is shown in equation (3).   
\item
  it is an interesting feature of the suggested approach that the position of the focused image of the reflection point P can be determined without knowledge of the true velocity model: after demigration it is necessary to estimate the residual one way radii of curvature at the shot and receiver positions which can be obtained in principle from diffraction analyses of shot/receiver surveys. In practice suitable approximations about the true velocity model should be used.  
\end{itemize}

\section{Applications}
\begin{figure}
  \captionsetup{justification=justified, singlelinecheck=off}
\centering
\includegraphics[width=0.45\textwidth]{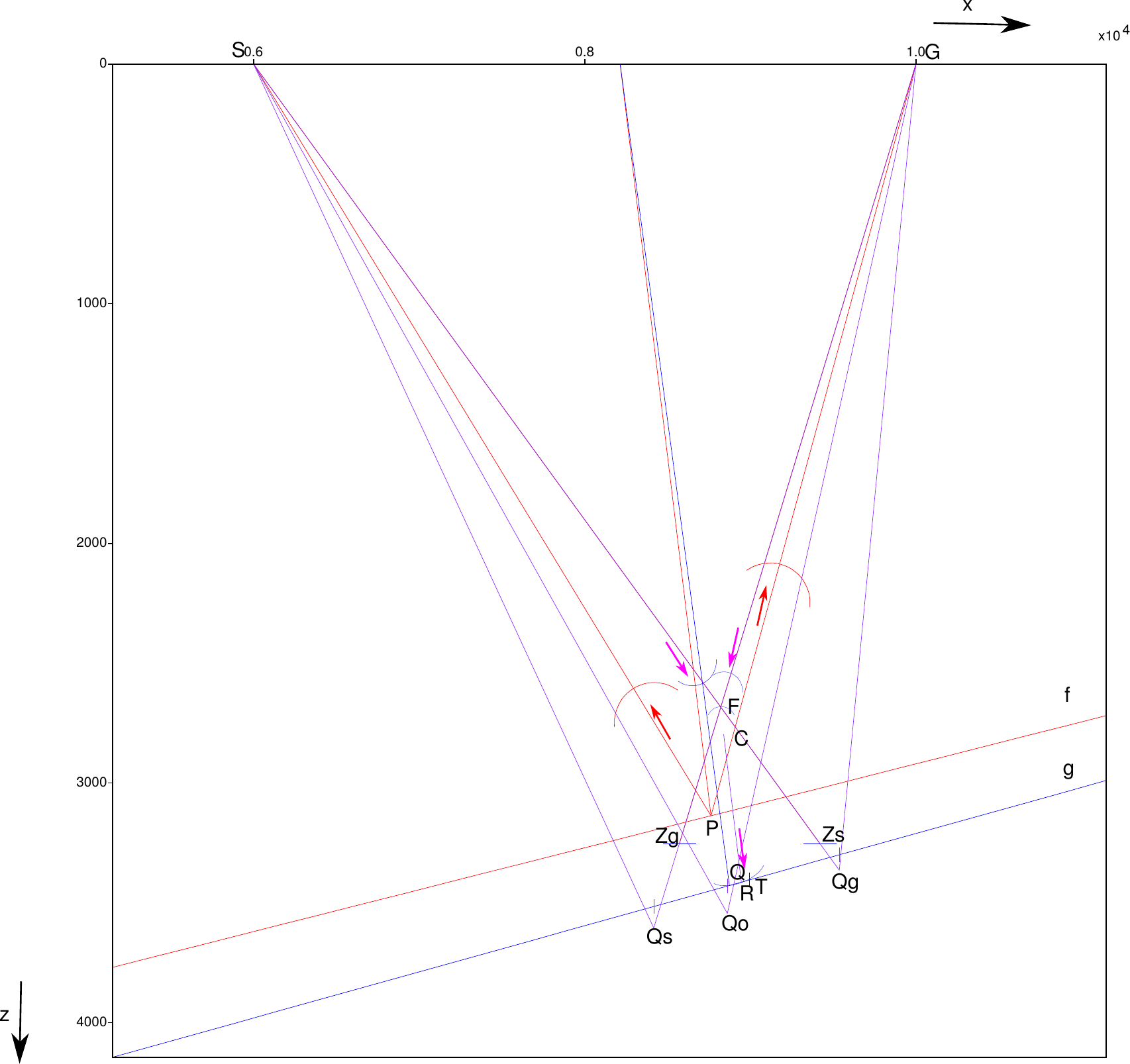}
\caption{Approximate RPSM to zero offset}
\label{fig:fig3}
\end{figure}

\begin{figure}
  \captionsetup{justification=justified, singlelinecheck=off}
\centering
\includegraphics[width=0.45\textwidth]{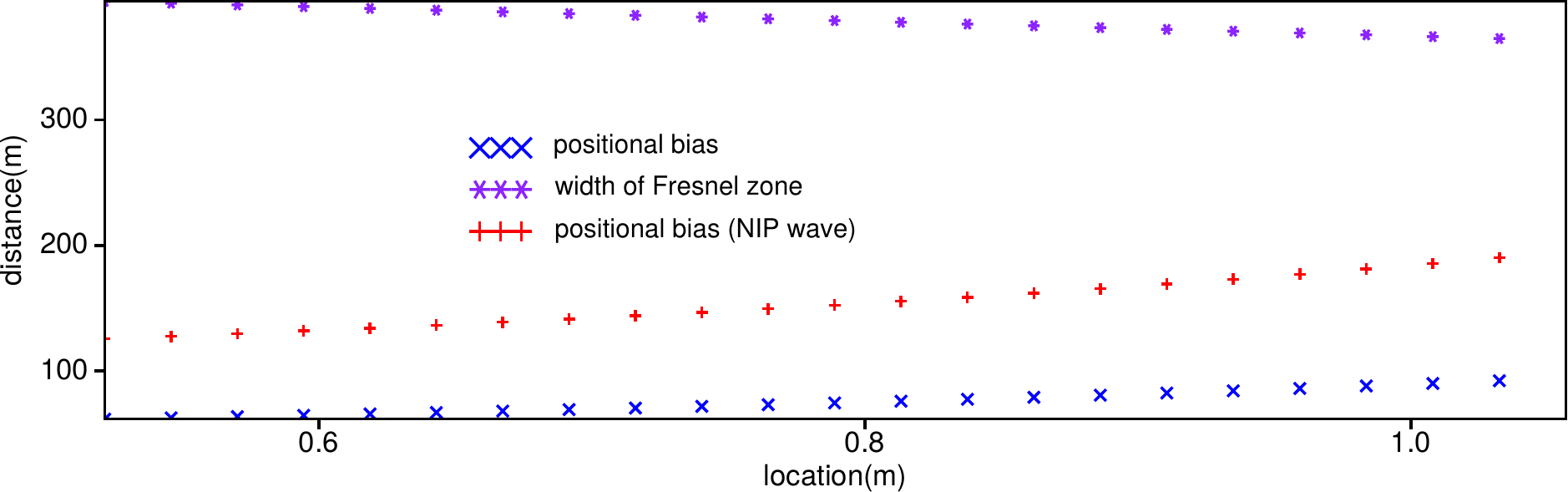}
\caption{Positional bias and width of Fresnel apertures}
\label{fig:fig4}
\end{figure}

In Figure~\ref{fig:fig3} a different configuration with opposite dip and $v_M > v_T$ is presented at a particular surface source-receiver offset of h=4000m (the computation were performed with true dip angle tan($\gamma$)=-0.175, $v_t$=2600m/s, $v_m$=2800m/s). Also shown are the results of (Kirchhoff) common offset, common shot and common receiver migrations at $Q_o$, $Q_s$ and $Q_g$ and their respective  zero offset positions (not labeled). It is apparent that the common shot and common receiver migrations both exhibit a large positional bias, whereas the common offset migration $Q_o$ is laterally close to to the image of the zero offset of the reflected event ray at Q.\\
Figure~\ref{fig:fig4} shows the half width of the positional bias both for the NIP wave approximation and for the suggested approach as well as the width of the first order Fresnel zone for a surface source receiver offset of 4000m. It is seen that the positional bias of the NIP wave approximation is considerably larger than for the suggested approach. The width of the Fresnel zone varies between 18 and 20 traces (with an CIG spacing of 20m: for practical applications the width of the zone should be confined between thresholds, depending on the residual moveout differences determined).\\

\begin{figure}
\captionsetup{justification=justified, singlelinecheck=off}
\centering
\includegraphics[width=0.45\textwidth]{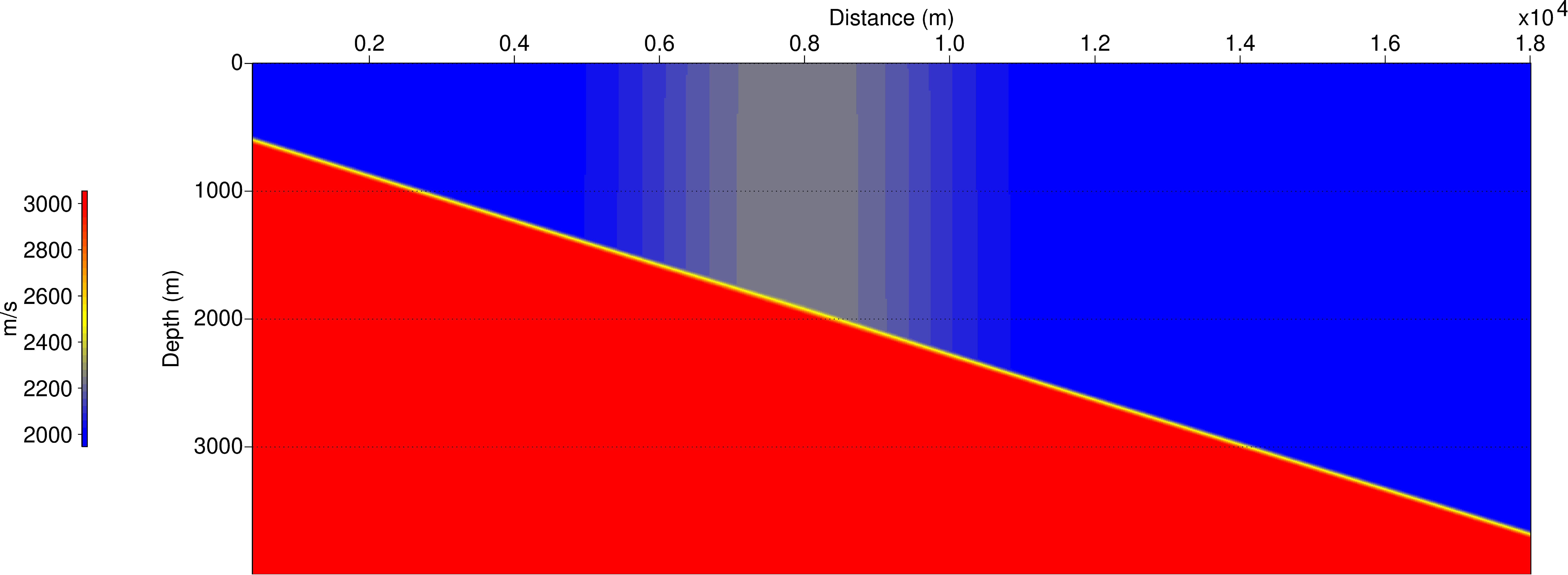}
\caption{Velocity model}
\label{fig:fig5}
\end{figure}

\begin{figure}
\captionsetup{justification=justified, singlelinecheck=off}
\centering
\includegraphics[width=0.45\textwidth]{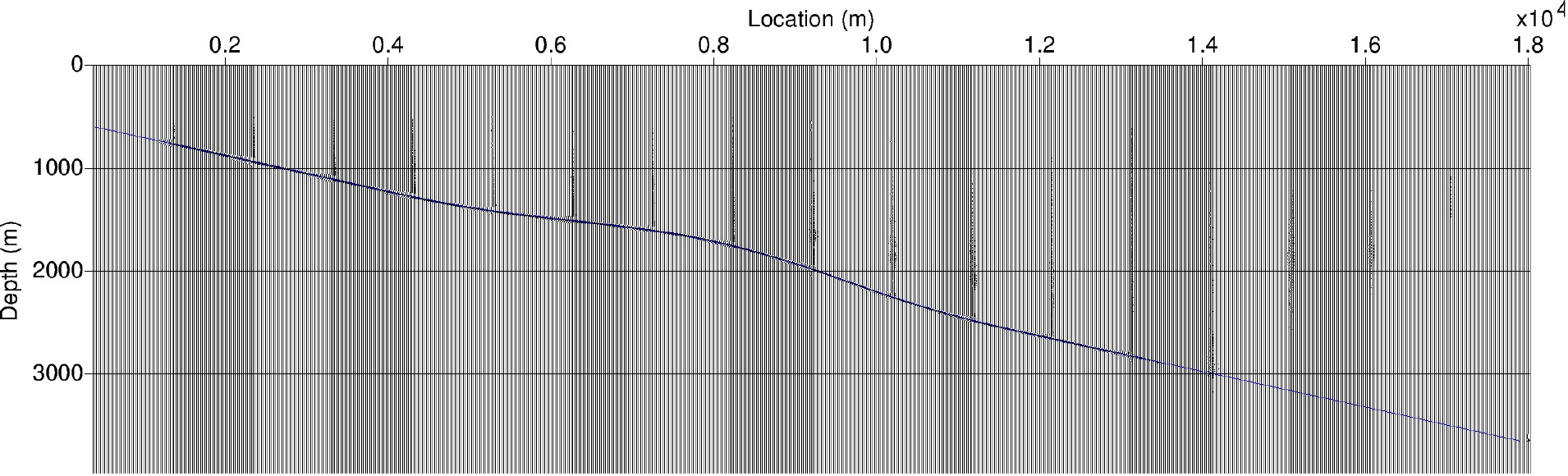}
\caption{Approximate RPSM with computed position indicated}
\label{fig:fig6}
\end{figure}

\begin{figure}
  \captionsetup{justification=justified, singlelinecheck=off}
\centering
\includegraphics[width=0.45\textwidth]{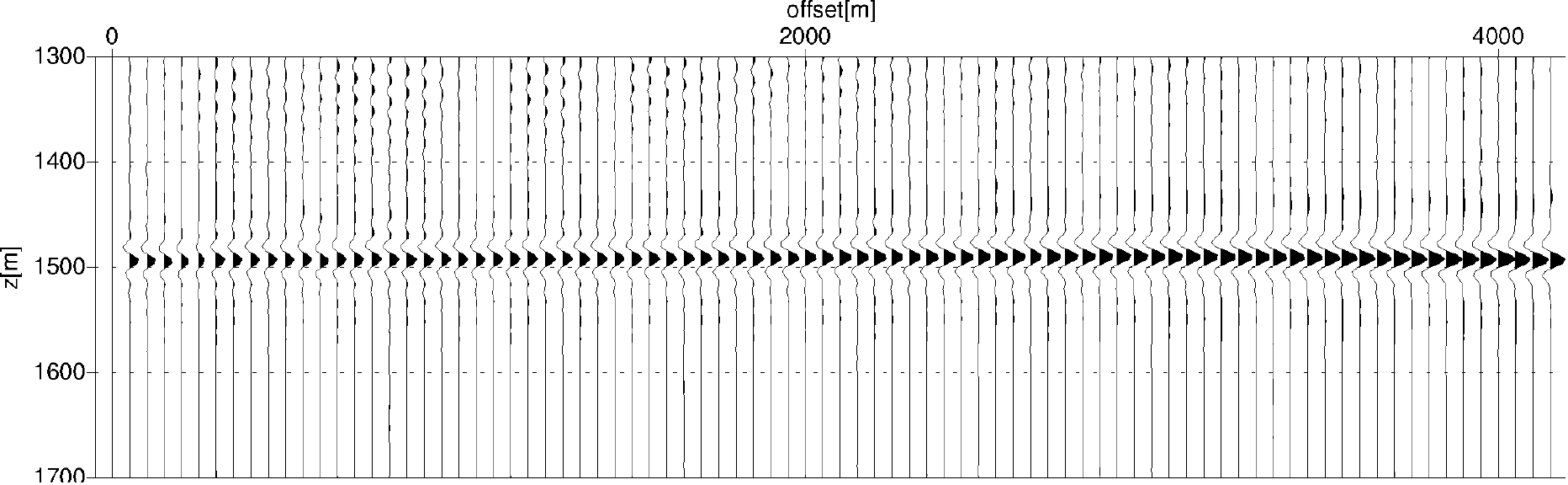}
\caption{CIG at 7850m after approximate RPSM}
\label{fig:fig7}
\end{figure}

\begin{figure}
  \captionsetup{justification=justified, singlelinecheck=off}
\includegraphics[width=0.45\textwidth]{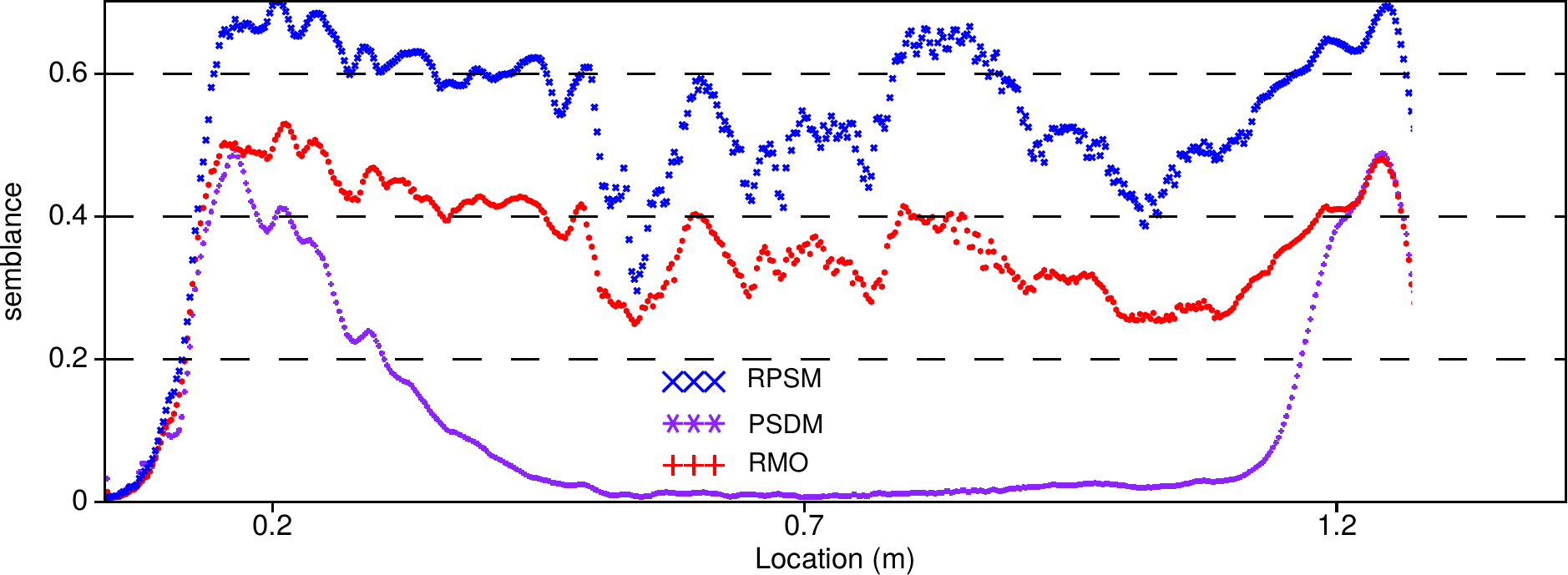}
\caption{Semblance, PSDM, after RMO and after RPSM}
\label{fig:fig8}
\end{figure}

The velocity model in Figure~\ref{fig:fig5} features two layers with a bell shaped velocity inhomogeneity in the overburden as in (Schneider, 2017). The stack of the suggested approximate RPSM in Figure~\ref{fig:fig6} coincides with the computed position of the migrated position (shown in blue), where the PSDM was performed with a constant velocity. The semblances in Figure~\ref{fig:fig8} are low for the migrated stack because of the incorrect migration velocities (Figure~\ref{fig:fig6}). After RMO corresponding values are still low due to a relatively large amount of migration noise (individual gathers are well corrected after RMO, Figure~\ref{fig:fig7}). The application of the approximate RPSDM is quite similar but features higher semblances, mainly due to the Fresnel summation.\\

\begin{figure}
\captionsetup{justification=justified, singlelinecheck=off}
\centering
\includegraphics[width=0.45\textwidth]{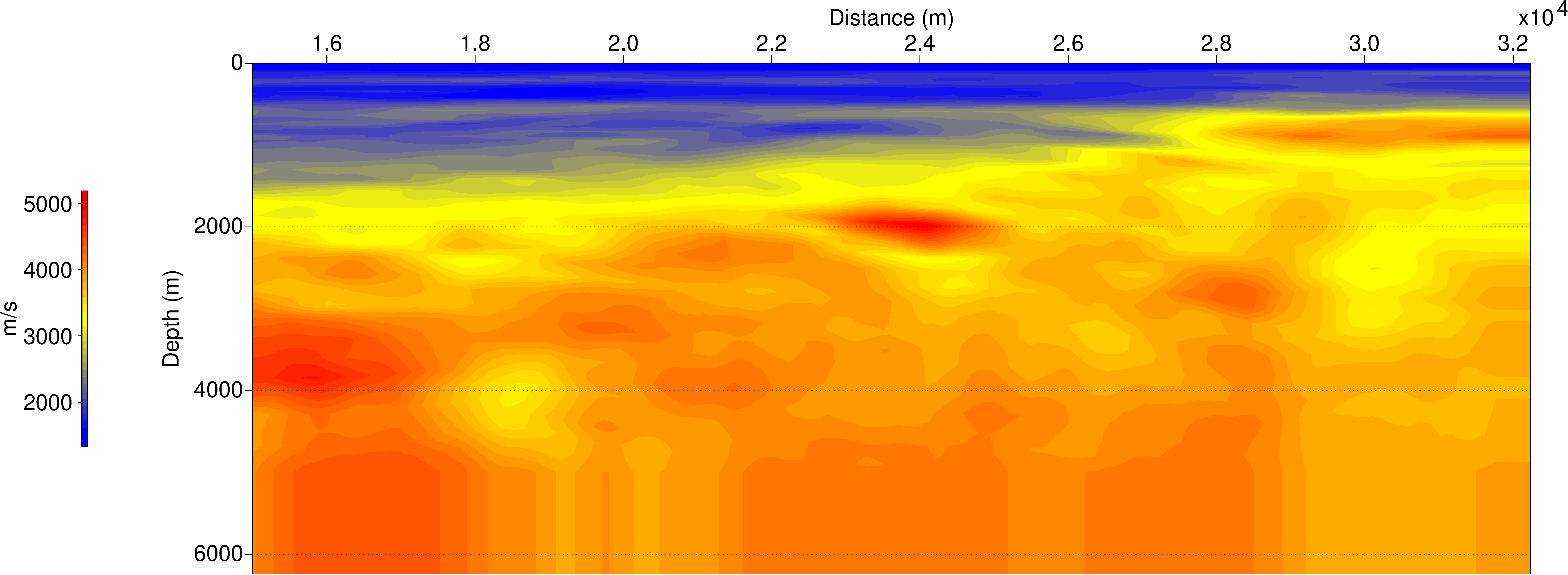}
\caption{Velocity model for PSDM}
\label{fig:fig9}
\end{figure}

\begin{figure}
  \captionsetup{justification=justified, singlelinecheck=off}
\centering
\includegraphics[width=0.45\textwidth]{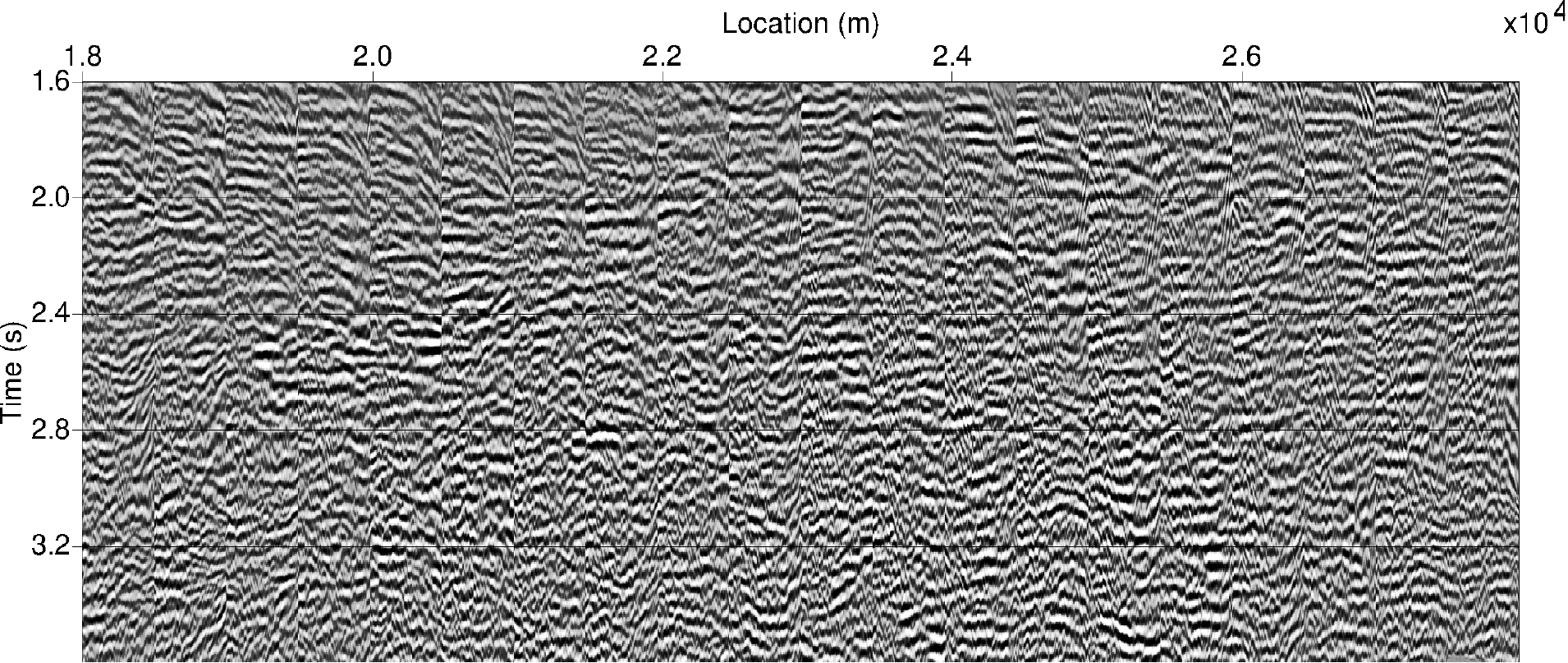}
\caption{CIG after PSDM and RMO (time scale)}
\label{fig:fig10}
\end{figure}

\begin{figure}
  \captionsetup{justification=justified, singlelinecheck=off}
\centering
\includegraphics[width=0.45\textwidth]{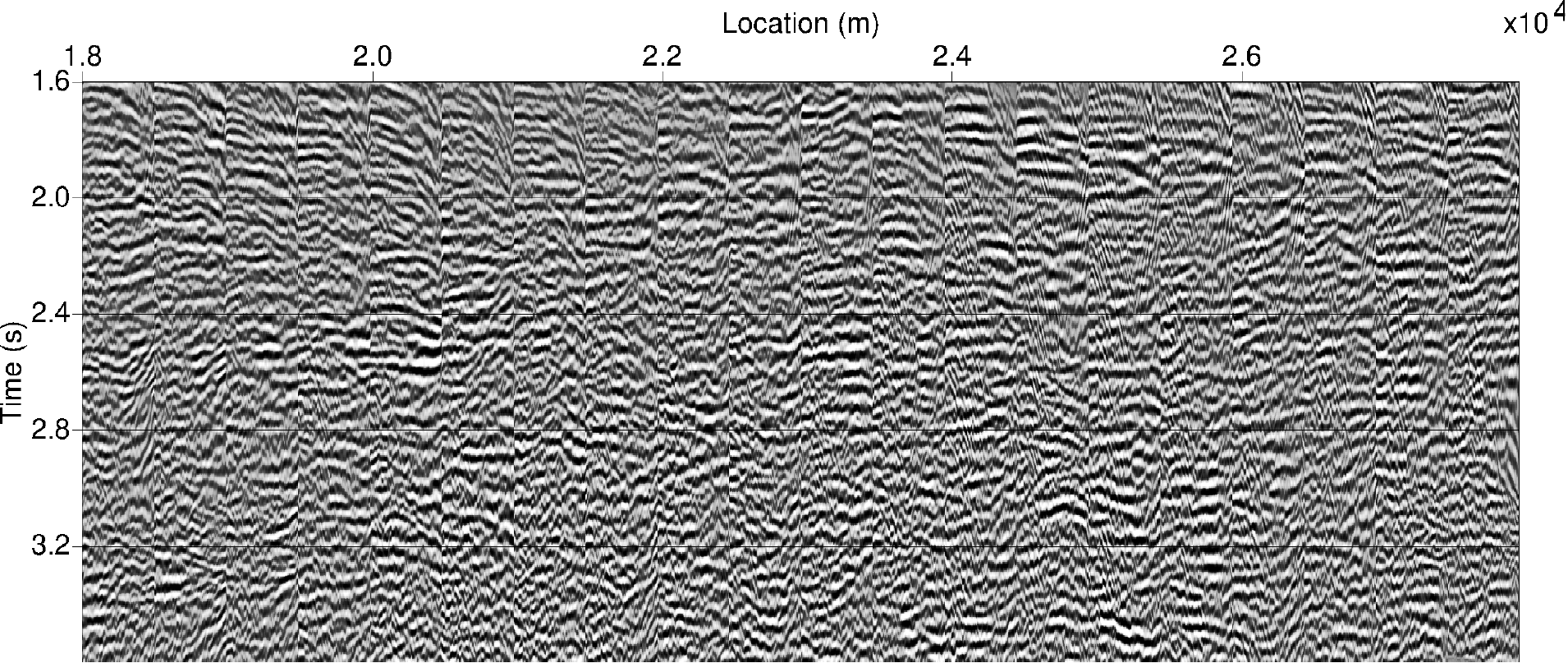}
\caption{CIG after PSDM and approximate RPSM to zero offset}
\label{fig:fig11}
\end{figure}

\begin{figure}
  \captionsetup{justification=justified, singlelinecheck=off}
\centering
\includegraphics[width=0.45\textwidth]{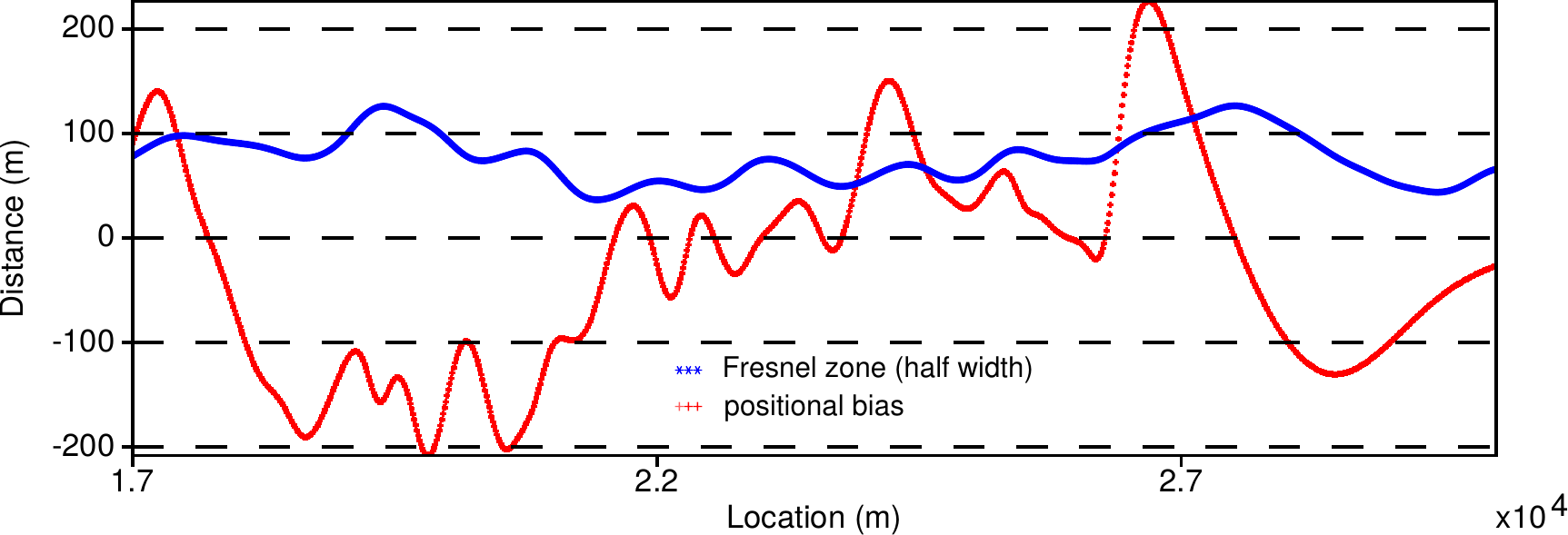}
\caption{Positional bias and half width of Fresnel zone}
\label{fig:fig12}
\end{figure}

\begin{figure}
  \captionsetup{justification=justified, singlelinecheck=off}
\centering
\includegraphics[width=0.45\textwidth]{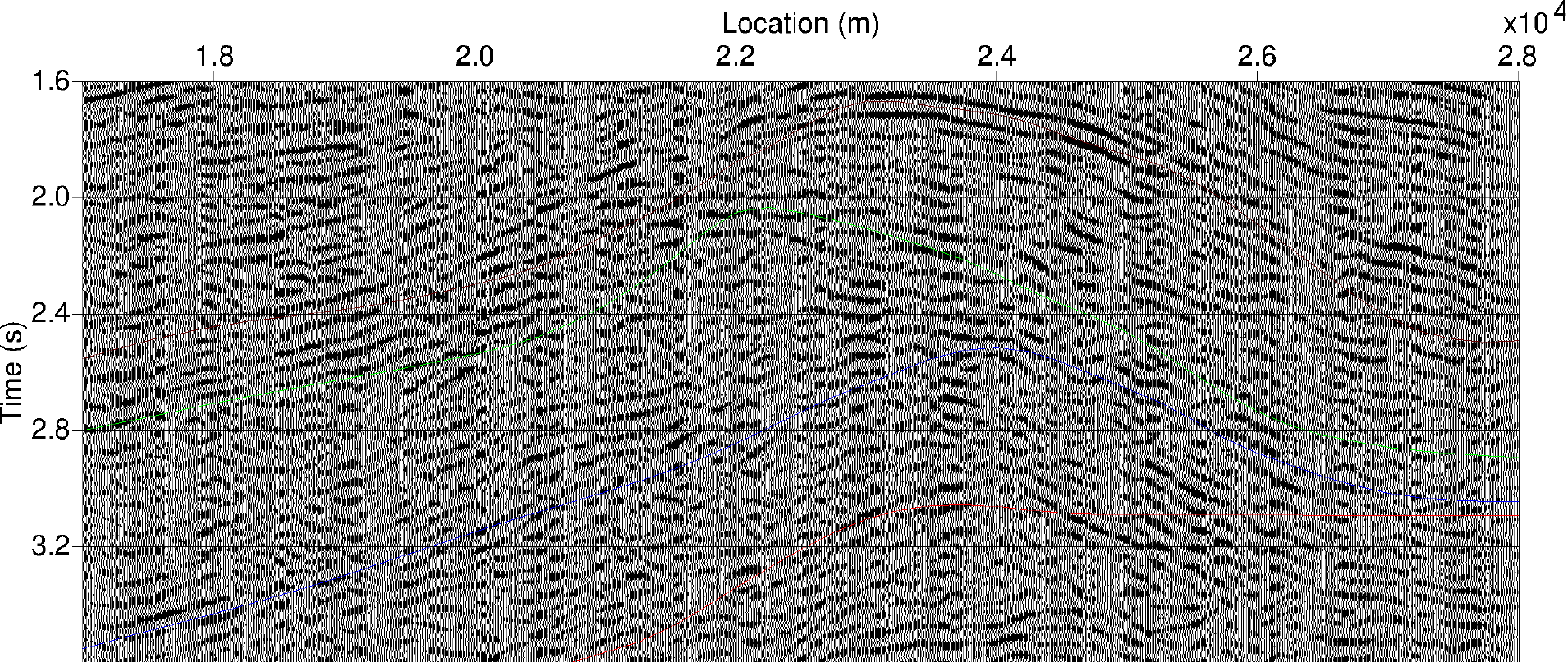}
\caption{PSDM and RMO (time scale)}
\label{fig:fig13}
\end{figure}

\begin{figure}
  \captionsetup{justification=justified, singlelinecheck=off}
\centering
\includegraphics[width=0.45\textwidth]{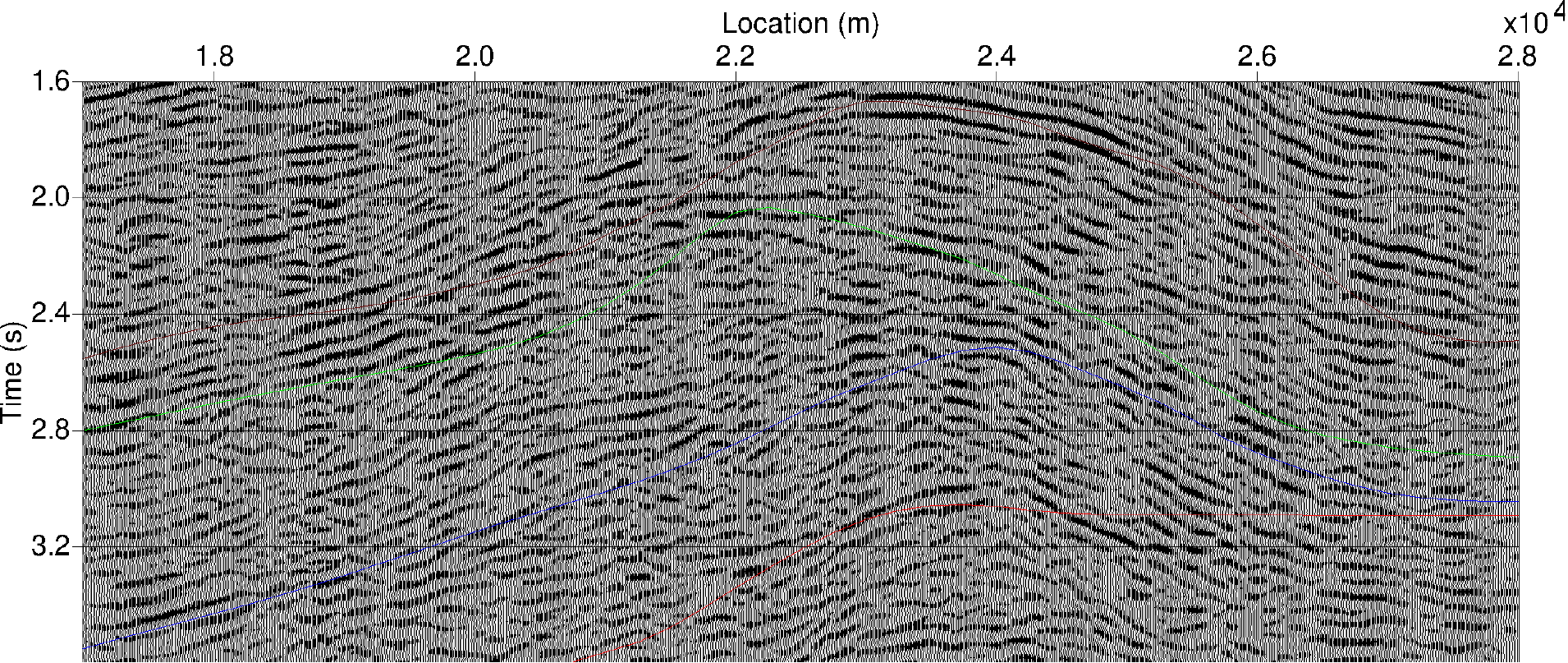}
\caption{Approximate RPSM to zero offset}
\label{fig:fig14}
\end{figure}

Figure~\ref{fig:fig9} shows a complicated velocity model; the centre of the section features an overthrust structure in a region of active exploration interest as in (Schneider, 2019). The PSDM of the free seismic survey (maximum offset 3200m, 120-fold) shows a relatively poor signal to noise ratio for the CIG (Figure~\ref{fig:fig10}, every 40th gather displayed) and the migrated stack (Figure~\ref{fig:fig13}, displayed in two way times, according to the velocity model). The RMO for the PSDM and the determination of the parameters for the approximate RPSM were performed along horizons as indicated in Figure~\ref{fig:fig13} (see Figure~\ref{fig:fig12} for one horizon). An application of the suggested application in Figures~\ref{fig:fig11} and ~\ref{fig:fig14} is shown here to demonstrate important features of the approach: the application of the Fresnel zone summation has clearly increased the S/N ratio. In the central and bottom part discontinuities are revealed which possibly indicate fault positions.\\

\section{Conclusion}

Travel times of reflected events for individual seismic traces have been determined for diffraction points situated along horizons of analysis and were downward continued into the migration model under consideration. The NIP wave theorem was applied to events exhibiting small subsurface offsets in order to determine residual radii of curvature which could be used in the updating of seismic velocity model.\\

An approach for an approximate residual migration to zero offset has been suggested for the case that a PSDM followed by an RMO analysis had been performed for a seismic survey. For the RPSM aplanatic curves were determined for individual horizons over the first Fresnel zone for the velocity model used for the migration. A summation was performed where parameters were obtained from interpolation between adjacent horizons. Amplitude and phase corrections were included to balance the distortions produced by the summation.\\

The approximate RPSM to zero offset was applied to synthetic data and to a seismic survey over an overthrust structure, in both cases after PSDM and RMO analyses. An increase in signal to noise ratio was observed for the migrated gathers and the migrated stack of the seismic survey whereas the lateral resolution was decreased in a controlled way.\\
\section*{Acknowledgments}

The author wishes to thank Dr. Sven Treitel for longstanding advice and encouragement and the provider of the seismic survey for preliminary processing. The Seismic Unix package has been used for most of the processing and graphical presentations.

\section{references}
Al-Chalabi, M., 2014, Principles of seismic velocities and time-to-depth conversion: EAGE.\\
Bednar J., 2005, A brief history of seismic migration: Geophysics, 70, 3MJ-20MJ.\\
Bleistein, N., Cohen, J., and F. Hagin, 1987, Two and one-half dimensional Born inversion with an arbitrary reference, Geophysics, 52, 26-32.\\
Born, M., 1985, Optik: Springer.\\
Dafni, M. and R. Symes, 2018: Kinematics of reflections in subsurface offset and angle-domainimage gathers, Geophys. J. Int., 213, 1212–123.\\
Douma, M. and de Hoop, H., 2006, Explicit expressions for prestack map time migration in isotropic and VTI media and the applicability of map depth migration in heterogeneous anisotropic media: Geophysics, 71, 513-528.\\
Guillaume, P., Lambare, G., Leblanc, O. et alii ,2008, Kinematics invariants: an efficient and flexible approach for velocity model building: 78th SEG Annual International Meeting, Expanded Abstracts.\\
Hagedoorn, J., 1954, A process of seismic reflection interpretation: Geophysical Prospecting, 2, 85-127.\\
Hubral, P. and Krey, T., 1980, Interval velocities from seismic refection time measurements: SEG .\\
Lambare, G., Herrmann, P., Toure, J., Suaudeau, E. and D. Lecerf, 2008, Computation of kinematic attributes for prestack time migration: SEG, 78th Annual annual meeting, Expanded Abstracts, 2402-2406.\\
Lindsey, J.P. 1989, The Fresnel zone and its interpretative significance: The Leading Edge, 8, 33-39.\\
Robein, E., 2010, Seismic Imaging: EAGE.\\
Schneider, J., 1989, Specular prestack migration, 51th EAGE conference\\
Schneider, J., 2017, Kinematic aspects of residual prestack migration: SEG, 87th Annual international meeting, Expanded Abstracts.\\
Schneider, J., 2019, Horizon oriented residual prestack migration to zero offset, SEG, 89th Annual international meeting, Expanded Abstracts.\\
Schneider, J., 2020, Kinematic aspects of residual prestack migration: arXiv:2003.12888 [physics.geo-ph]\\
Sheriff R.E. 1980. Nomogram for Fresnel zone calculation: Geophysics, 45, 968-972.\\
Yilmaz, O, 2001, Seismic Data Analysis: SEG.\\

\end{document}